\documentclass{kluwer}
\usepackage{epsfig}
\begin{document}
\begin{opening}
\title{Parallel TreeSPH: A Tool for Galaxy Formation}
\author{C. \surname{Lia}\email{liac@sissa.it}}
\institute{SISSA/ISAS, via Beirut 2, I-34013 Trieste, Italy}
\author{G. \surname{ Carraro}\email{carraro@pd.astro.it}}
\institute{Department of Astronomy, Padova Univesity, Vicolo
Osservatorio  5, I-35122, Padova, Italy}
\begin{abstract}
We describe a new implementation of a parallel Tree-SPH code with 
the aim to simulate Galaxy Formation and Evolution.
The code has been parallelized using SHMEM, a Cray proprietary library
to handle communications between the 256 processors of the Silicon 
Graphics
T3E massively parallel supercomputer hosted by the Cineca
Super-computing Center (Bologna, Italy).\\   
The code combines the Smoothed Particle Hydrodynamics (SPH) method
to solve hydro-dynamical equations
with the popular Barnes and Hut (1986) tree-code to perform gravity
calculation with a $N \times logN$ scaling, and it is based on the  
scalar Tree-SPH
code developed by Carraro et al (1998)[MNRAS 297, 1021].\\
Parallelization is achieved distributing particles along processors
according to a work-load criterium.\\
Benchmarks, in terms of load-balance and scalability, of the code are
analised and critically discussed against the
adiabatic collapse of an isothermal gas sphere test using $2 \times
10^{4}$
particles on 8 processors.
The code
results balanced at more than $95\%$ level. Increasing the number of
processors, the load balance sligthly worsens.
The deviation from perfect
scalability at increasing number of processors is negligible up to 
64 processors.
Additionally we have incorporated radiative cooling, star formation,
feed-back and an algorithm to follow the chemical enrichment of the
interstellar medium. 
\end{abstract}
\end{opening}
\section{Introduction}
Carraro et al (1998) developed a pure particle code, combining Barnes \& Hut (1986) octo-tree
with SPH,  and applying this code to the formation of a spiral galaxy like the Milky Way.
The code is similar to Herniquist \& Katz (1989) TreeSPH. It uses SPH to solve the
hydro-dynamical equations. In SPH a fluid is sampled using particles, there is no resolution
limitation due to the absence of grids, and great flexibility thanks to the use of a time and space
dependent smoothing length.
Shocks are captured by adopting an artificial viscosity
tensor,  and
the neighbors search is performed using the octo-tree. The octo-tree, combined with SPH,
allows a time scaling  of $N\times logN$.  A good advantage of such codes is that it is easy to
introduce new physics, like cooling and radiative processes, magnetic fields and so forth.
Finally the kernel, which is utilised to performe hydro-dynamical
quantities estimates,  can be made adaptive by using anisotropic smoothing lengths.
It is widely recognized that TreeSPH codes, although deficient
in some aspects,  can give reasonable answers in many astrophysical situations, like in simulations
of fragmentation and star formation in giant molecular clouds (GMC)(), supenov\ae explosions
globular clusters formation,  merging of galaxies, galaxies
and clusters formation 
and  Lyman alpha forest.
Galaxy formation in particular requires a huge dynamical range (Dav\'e et al 1997). In fact
an ideal galaxy formation simulation would start from a volume as large as the universe to   
follow
the initial growth of the cosmic structures, and  at the same time would be able to resolve
regions as small as GMC, where stars form and drive the galaxy evolution through their
interaction with ISM. This ideal simulation would encompass a dynamic range of $10^{9}$
(from Gpc to parsec), $10^{6}$ time smaller than that achievable with present day codes.
Big efforts have been made in the last years to enlarge as much as possible the dynamical range
of  numerical simulations, mainly using more and more powerful supercomputers.
Scalar and vector computers indeed cannot handle efficiently a number of particles greater than
half a million .
Dav\'e  et al (1997) for the first time developed a parallel implementation of a TreeSPH
code (PTreeSPH) which can follow both collision-less and collisional matter.
They report  results of simulations run on a Cray T3D computer of the adiabatic collapse
of an initially isothermal gas sphere (using 4096 particles),  of  the collapse of a
Zel'dovich pancake (32768 particles) and of a cosmological simulation
(32768 gas and 32768 dark particles).
Their result are quit encouraging, being quite similar to those obtained with the scalar
TreeSPH code (Hernquist \& Katz 1989).
Porting a scalar code to a parallel machine is far from being an easy task.  A massively
parallel computer  (like the Silicon Graphics T3E) links together hundreds
or thousands of processors
aiming at increasing  significantly the computational power. For this reason they are very
attractive, although several  difficulties can  arise in adapting a code to these machines.
Any processor possesses its own memory, and can assess other processors memory by
means of communications which are handled by an hard-ware network, and are slower than
the computational speed.
Great attention must be paid to avoid situations in which a small number of processors
are actually working while most of them are standing idle. Usually one has to invent
a proper data distribution scheme which allows to subdivide particles into processors
in a way that any processor handles about the same number of particles and does not
need to make heavy communications. Moreover the computational load must be
shared between processors, ensuring that  processors exchange informations
all together, in a synchronous way, or that  any  processors is performing different
kinds of work  when it is waiting for informations coming from other processors, in an
asynchronous fashion (Dav\'e et al 1997).
In this paper we present a parallel implementation of the TreeSPH code dscribed in Carraro
et al (1998). The numerical ingredients are the same as in the scalar version of the code.
However the design of the parallel implementations required several changes to the
original code.
The key idea that guided us in building the parallel code was to avoid continuous
communications, limiting the informations exchange at a precise moment along the code flow.
This clearly reduces the communication overheard.
We have also decided to tailor the code to the machine, improving its efficiency.
Since we are using a T3E massively parallel computer, a natural choise was to handle
communications using the SHMEN libraries,  which permits asynchronous communications,
and are intrinsically very fast, being produced  directly by Cray for the T3E super-computer.
At present the code is also portable to other machine, like SGI Origin 2000, and will be
portable to any other machine with the advent of the second release of  Message Passing Interface
(MPI).

\begin{figure*}
\centerline{\epsfig{file=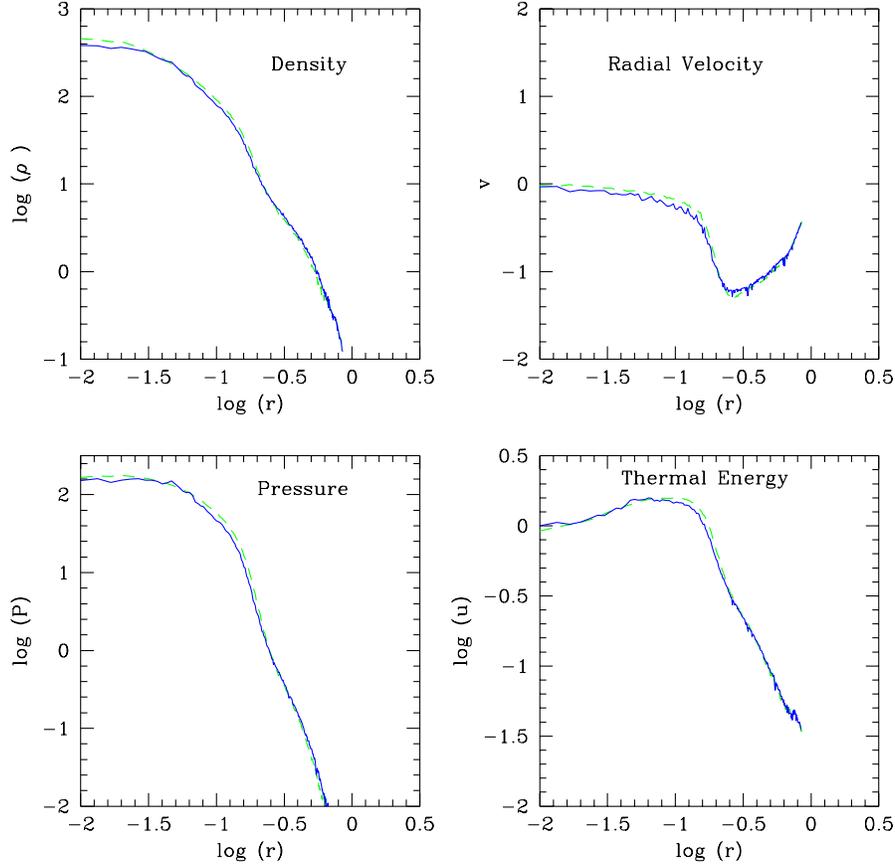,width=12cm}}
\caption{Adiabatic collapse: snapshots of the density, radial velocity,  
pressure and internal energy at the time of the maximum compression.
The results of the test
performed using $2\times 10^{4}$ particles with 8 processors are shown
with dashed lines. Solid lines show the results obtained with the same
number of particles, but using the scalar
code, for comparison.}
\end{figure*}

\section{A test of the code}
We consider the adiabatic collapse of an
initially non-rotating
isothermal gas sphere. This is a standard test for  SPH codes
(Hernquist \& Katz 1989.
In particular it is an ideal test for a parallel code,
due to the large dynamical range and high density contrast.
To facilitate the  comparison of  our results with those by the
above authors, we adopt the same
initial model and the same units ($M=R=G=1$).
The system consists of a $\gamma = 5/3$ gas sphere, with an
initially isothermal density profile:
\begin{equation}
\rho(r) = \frac{M(R)}{2\pi R^{2}} \frac{1}{r}  ,
\end{equation}
\noindent
where M(R) is the total mass inside the sphere of radius R.
The density profile is obtained stretching an
initially regular cubic grid.

\begin{figure}
\centerline{\epsfig{file=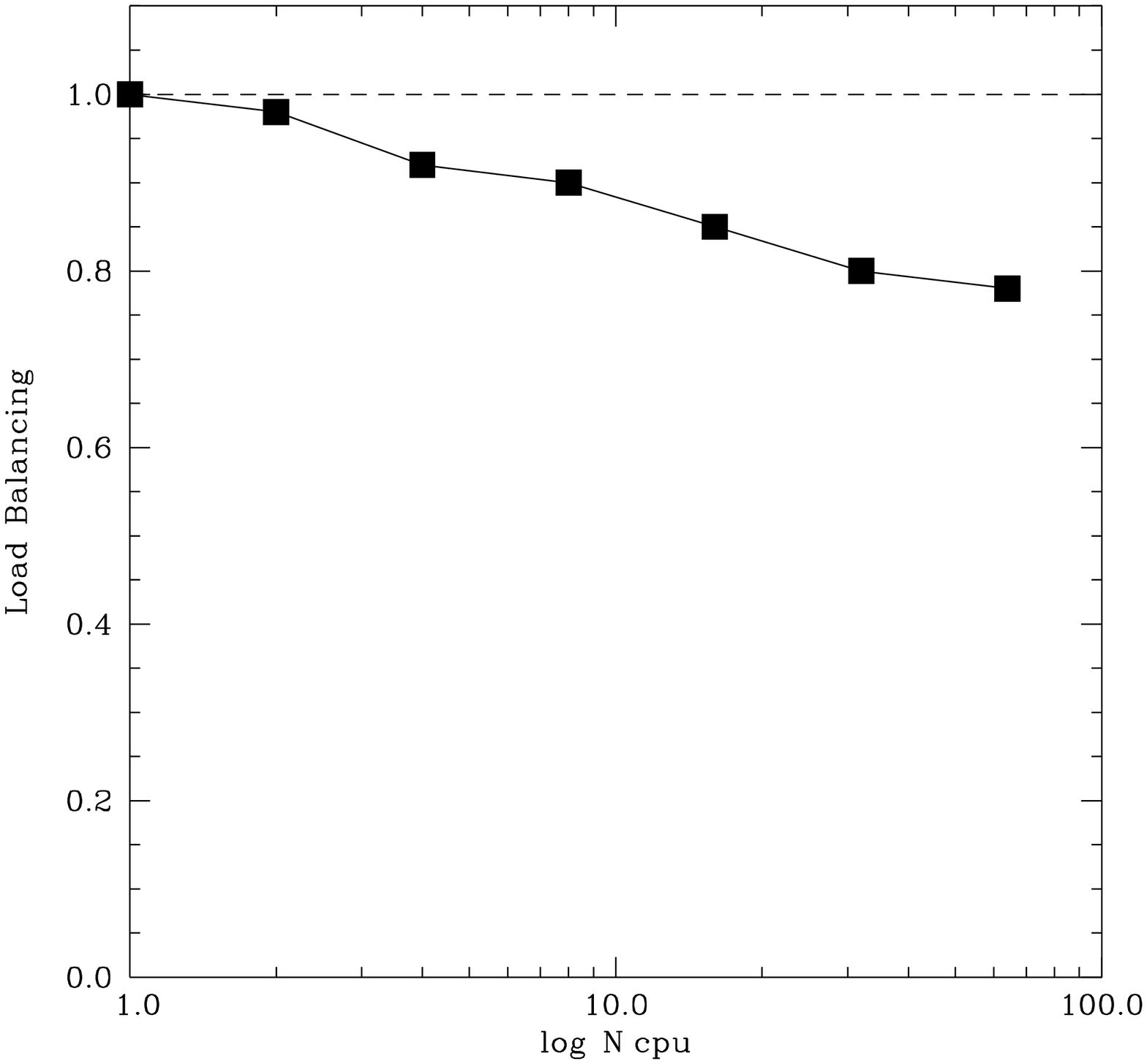,width=8cm}}
\caption{Overall code load-balance, averaged on 50 time-steps (solid  
line).
Dashed line indicates ideal scalability.}
\end{figure}

\begin{figure}   
\centerline{\epsfig{file=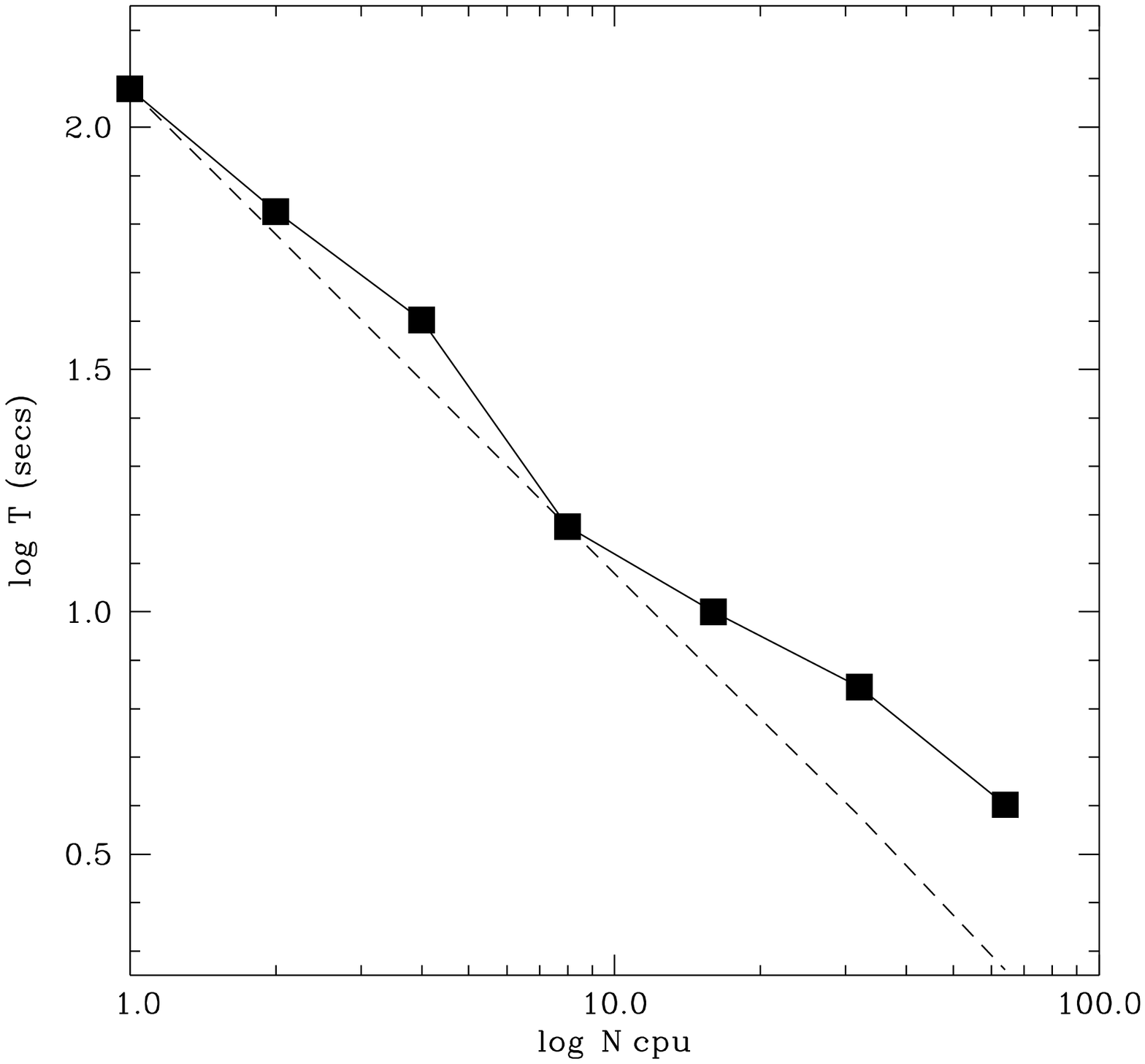,width=8cm}}
\caption{Overall code scalability, averaged on 50 time-steps (solid line).
Dashed line indicates ideal load-balance.}
\end{figure}

The  total number of particles used in this simulation
is $2\times 10^{4}$. All the particles have the same mass. 
The specific internal energy is set to $u = 0.05GM/R$.
For this test the viscosity parameters $\alpha$ and $\beta$
adopted are 0.5, in agreement with Dav\'e et a (1997).
The gravitational softening parameter $\epsilon$ adopted for
this simulation is $5 \times 10^{-3}$.
The state of the system at the time of the maximum compression
is shown in the various panels  of Fig. 1,
which displays the density, radial velocity, pressure  and specific internal
energy profiles. Each panel shows the variation of the physical
quantity under consideration (in suitable units) as a function of the
normalized radial coordinate at time equal to 0.88 .   
The initial low internal energy is not sufficient to support the gas
cloud which starts to collapse. Approximately after
one dynamical time scale a bounce
occurs. The system afterwards can be described as an isothermal core plus an
adiabaticlly expanding envelope
pushed by the shock wave generated at the stage of
maximum compression.
After about three dynamical times the system reaches virial equilibrium
with total energy equal to a half of the gravitational potential energy
(Hernquist \& Katz 1989).
The present results agree
fairly well with the mean values of the Hernquist \& Katz (1989)
simulations, which in turns agrees with the 1-D finite difference
results. 

\begin{figure}
\centerline{\epsfig{file=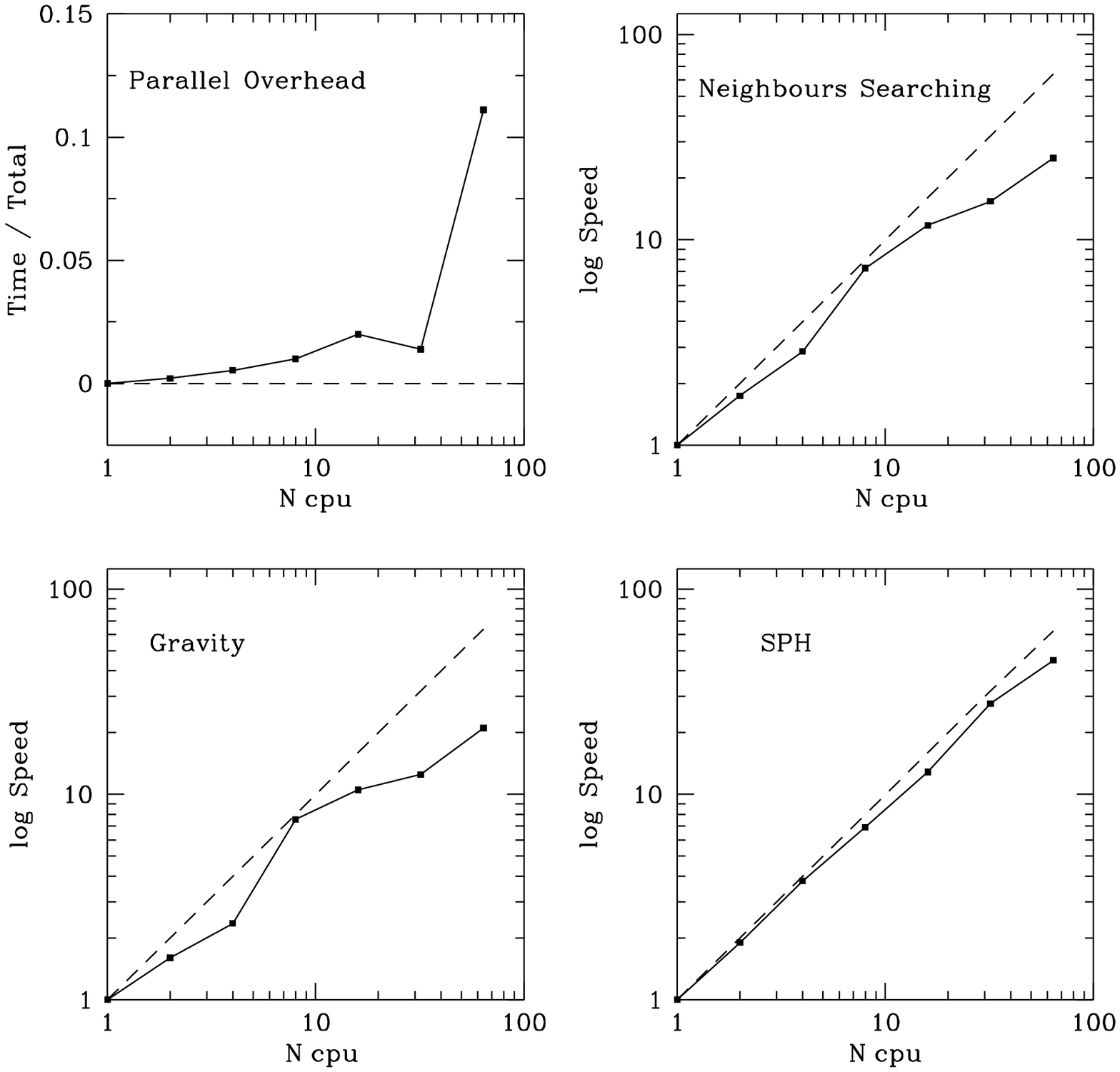,width=12cm}}
\caption{Code scalability in different code sections, averaged on 50 time-steps (solid line).
Dashed line indicates ideal load-balance.}
\end{figure}

\begin{table*}
\tabcolsep 0.1truecm
\caption{The Abiabatic Collapse test.
Bechmarks for a run with $2\times 10^{4}$ particles. Time refers
to 50 time-stpes.}
\begin{tabular}{cccccccc} \hline
\multicolumn{1}{c}{$N_{cpu}$} &
\multicolumn{1}{c}{Total} &
\multicolumn{1}{c}{Data Up-date} &
\multicolumn{1}{c}{Parallel Over-head} &
\multicolumn{1}{c}{Neighbours} &
\multicolumn{1}{c}{SPH} &
\multicolumn{1}{c}{Gravity} &
\multicolumn{1}{c}{Miscellaneous} \\
\hline
& secs &secs & secs & secs & secs & secs & secs\\
\hline
 1  &120    &0.47  &0.00    &40    &36    &40   &3.53\\
 2  &69     &0.22  &0.60    &23    &19    &25   &1.18\\
 4  &42     &0.27  &1.70    &14    &9.5   &15   &1.53\\
 8  &23     &0.13  &3.20    &5.5   &5.4   &5.3  &3.47\\
16  &17.3   &0.13  &3.40    &3.4   &3.0   &3.8  &3.60\\
32  &11.5   &0.09  &3.00    &2.6   &1.3   &3.2  &1.31\\
64  &7.5    &0.05  &2.90    &0.33  &1.1   &1.9  &1.23\\
\hline
\hline
\end{tabular}
\end{table*}

We run the adiabatic collapse test up to the time of the maximum compression
(t $\simeq 1.1$) using $2 \times 10^{4}$ particles
on 1, 2, 4, 8, 16, 32 and 64 processors, and looked at the performances in
the following code sections (see also Table~1):

\begin{description}
\item[$\bullet$] total wall-clock time;
\item[$\bullet$] data up-dating data;
\item[$\bullet$] parallel computation,
which consists of barriers, the construction of the {\it ghost-tree} and the
distribution of data between processors;
\item[$\bullet$] search for neighbour particles;
\item[$\bullet$] evaluation of the hydro-dynamical quantities;
\item[$\bullet$] evaluation of the gravitational forces;
\item[$\bullet$] miscellaneous, which encompasses I/O and kernel computation.
\end{description}

\noindent
The results summarized in Table~1 present the total
wall-clock time per timestep per processor, averaged over 50 time-steps,
together with the time spent in each of the 5 subroutines
(data updating, neighbour searching, SPH computation, gravitational
interaction and parallel computation).
The gravitation interaction takes about one-third of the total time,
while the search for neighbours takes roughly comparable time.
The evaluation of hydrodynamical quantities takes about
one-fourth of the time, the remaining time being divided between I/O and
data up-dating. The parallel over-head  does not appear to be a problem, being
always less than $1\%$ of the total time.
This timing refers, as indicated above, to simulations stopped at roughly the time
of maximum compression. A run with 8 processors up to $t \simeq 2.5$, the time at which
the system is almost completely virialized,
took 3800 secs.
One of the most stringest requirement for a parallel code is the capability to distribute
the computational work equally between all processors.
This can be done defining a suitable work-load criterium, as discussed in Section~3.2.
This is far from being an easy task (Dav\`e et al 1997), and in practice some
processors stand idly for some time waiting that the processors with the greatest
computational load accomplish their work. This is true also when an asynchronous communications
scheme is adopted, as in our TreeSPH code.
To evaluate the code load-balance we adopted the same strategy of Dav\`e et al (1997),
measuring the fractional amount of time spent idle in a time-step
while another processor performs computation:
\begin{equation}
L = \frac{1}{N_{procs}}\sum_{j=1}^{N_{procs}} 1 - \frac{(t_{max - t_i})}{t_{max}}   .
\end{equation}
Here $t_{max}$ is the time spent by the slowest processor, while $t_i$ is
the time taken by the $i-th$ processors to perform computation.
The results are shown in Fig~2, where we plot the load-balance for simulations
at increasing number of processors, from 1 to 64.
The load balance maintains always above $80\%$, being close to 1 up to 8 processors.
For the kind of simulation we are performing, the use of 8 processors is particulary  
advantageous for symmetry reasons.
At increasing number of processors, a parallel code should ideally speeds up linearly
In practice the increase of the processors number casuses an increase of the communications
between processors, and a degradation of the code performances.
To test this, we used the same simulations discussed above, ruuning the adiabatic collapse
test with $2 \times 10^{4}$ particles at increasing processors number.
We estimated how the code speed scales computing
the wall-clock time per processor
spent to execute a single time-step, averaged over 50 time-steps. In Fig.~3 we plot the
speed (in $sec^{-1}$ against the number of processors. \\   
The code scalabilty keeps very close to the ideal scalability up to 8 processors,
where it shows a minimum. This case in fact is the  most symmetricone.
Then the scalability deviates significantly only
when using more that 16 processors. Looking also at Fig.~4, it is easy to recognize that
mainly the gravitational interaction is responsible for this deviation.\\
In the near future we are going to investigate whether the code overall scalability
might improve adding new physics
(like cooling and star formation) which is necessary to describe the evolution of real systems,
like galaxies.

\end{document}